\documentclass{ws-procs9x6}
\usepackage{mathrsfs} 
\begin{document}

\title{Weighing the Dark Matter Halo}

\author{Jacob L. Bourjaily\footnote{{\tt jbourj@umich.edu}}}

\address{Michigan Center for Theoretical Physics, \\ University of Michigan, Ann Arbor, MI 48109-1120}

\maketitle

\abstracts{
\indent The dark matter problem will be solved only when all of the dark matter is accounted for. Although wimps may be discovered in direct detection experiments soon, we will not know what fraction of the dark matter halo they compose until we measure their local density. In this talk, I will offer a novel method to determine the mass of a wimp from direct detection experiments alone using kinematical consistency constraints. I will then describe a general method to estimate the local density of wimps using both dark matter detection and hadron collider data when it becomes available. These results were obtained in collaboration with Gordon Kane at the University of Michigan.}

\vspace{-0.8cm}
\section{Introduction}
The direct detection of wimps in the galactic halo would be an enormous triumph of experimental and theoretical particle cosmology, have deep implications for our understanding of the universe, and would explain (at least) some of the dark matter in the universe. However, it is unreasonable to assume that the entire dark matter halo is composed of only the wimp observed.\\
\indent What fraction of the dark matter halo is represented by a particular wimp is a question that cannot be answered by direct detection experiments alone or colliders alone. Even if a stable, weakly interacting massive particle is discovered at the LHC, no amount of collider data can measure its {\it actual} local density. Alternatively, even if many direct detection experiments unambiguously observe the scattering of wimps from the halo, these experiments cannot  identify the wimp or determine its couplings---crucial to measuring the local density. This point was first raised technically in Ref.\cite{Brhlik:2000dm}. In this talk, I will present a general way to address the cosmological significance of wimps observed in direct detection experiments using data and bounds from colliders. For a more detailed discussion, see Ref.\cite{Bourjaily:2004aj}.
\vspace{-.6cm}
\section{The No-Lose Theorem vs. Our Ability to Win}
\indent Let us imagine that a weakly interacting massive particle $\chi$ has been unambiguously observed in direct detection experiments. There is no reason to suspect that $\chi$ is {\it all} the dark matter. In fact, it is very easy to find models where detectable wimps compose only a fraction of a percent of all the dark matter. Using the framework of the DarkSUSY code\cite{Gondolo:2004sc}, we randomly generated some six thousand constrained MSSMs and found that for any signal rate, the relic density fluctuates over at least two orders of magnitude\cite{Bourjaily:2004aj}\footnote{Only an upper bound on the relic density, consistent with WMAP, was imposed while generating these models.}. It appears as if the detection signal rate and relic density are largely uncorrelated: a wimp with a low relic density may be just as observable as one making up most of the dark matter.\\
\indent The fact that particles with even small relic densities can have large detection signals has been noted by many authors (see, {\it e.g.} Ref.\cite{Duda:2002hf}) and has sometimes referred to as the `no-lose theorem:' experimentalists may not lose out on discovering even a very tiny faction of the dark matter halo. However, the no-lose theorem also implies that a wimp discovery could easily represent a negligible fraction of the dark matter. Therefore, although wimps may be discovered in the near future, the dark matter problem will not be solved until the density of wimps has been directly determined, and all the dark matter is accounted for.
\section{Dark Matter Direct Detection Rates}
\indent In general, the wimp-nucleon elastic scattering scattering rate is a function of the cross section for scattering, nuclear physics describing the detector, and the local velocity profile of the wimp fraction of the dark matter halo. If a detector is composed of nuclei labeled by the index $j$, each with mass fraction $c_j$, then the differential rate of wimp scattering at recoil energy $q$ is given by\footnote{A detailed discussion of equation \ref{rate} can be found in most modern reviews of dark matter, (see, {\it e.g.}, Ref.\cite{Jungman:1995df}).},
\begin{align}
\hspace{-0.85cm}\left.\frac{dR}{dQ}\right|_{Q=q}\!\!\!\!\!=&\frac{2\rho_{\chi}}{\pi m_{\chi}}\sum_{j}c_j\int_{v_{\mathrm{min}_j}(q)}^{\infty}{\!\!\!\frac{f(v,t)}{v}dv}\left\{\raisebox{0.5cm}{$\!$}\right.F_{j}^2(q)[Z_jf_p+(A_j-Z_j)f_n]^2\nonumber\\
&+\left.\frac{4\pi}{(2J_j+1)}\left[a_0^2S_{j_{00}}(q)+a_1^2S_{j_{11}}(q)+a_0a_1S_{j_{01}}(q)\right]\raisebox{0.5cm}{$\!$}\right\},\label{rate}
\end{align}
where $f(v,t)$ is the halo velocity profile, $F_j^2(q)$ and $S_{j_{mn}}(q)$ are nuclear form factors, $a_0\equiv a_p+a_n$ and $a_1\equiv a_p-a_n$, and the constant parameters $f_{p,n}$ and $a_{p,n}$ describe the coherent and incoherent wimp-nucleon scattering cross sections, respectively.\\
\indent From the equation above, it is clear that to determine the density $\rho_{\chi}$, one must identify the particle, determine its mass, estimate the halo profile, and `know' the interaction parameters from the theory describing $\chi$. Each of these require enormous efforts of both dark matter and collider experiments.
\vspace{-0.3cm}
\section{Determining the Wimp Mass}\begin{figure}[t]\vspace{-1cm}\centering\includegraphics[scale=0.81]{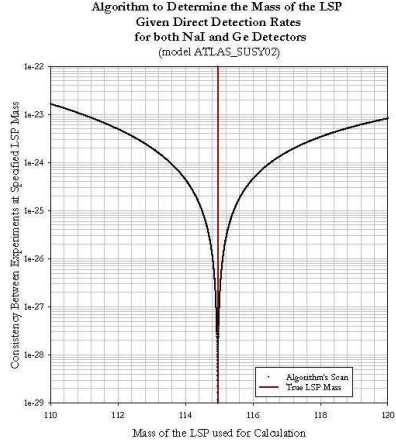} \caption{The function $\zeta(m'_{\chi})$ where the wimp corresponds to the neutralino in the MSSM specified by ATLAS SUSY point 2. The models and data were generated within the framework of the DarkSUSY package.}\label{mass_scan}\end{figure}
\indent Initially, perhaps the most important parameter of $\chi$ to determine is its mass. Not only does $m_{\chi}$ determine the particle's kinematics, but it may be critical to the identification of the particle\footnote{One may hope that accelerators observe a weakly interacting, stable, massive particle with the same mass as $\chi$.}.\\
\indent There are two known ways to determine $m_{\chi}$ from direct detection data alone. One method, using the annual modulation crossing energy, was described in a dark matter review article by Primack {\it et. al.} in 1988\cite{Primack:1988zm}\footnote{Although it seems unlikely to have originated in a review article, we have been unable to find any earlier reference.}. The other method has been developed by the author and is described presently.\\
\indent If the halo velocity profile and $m_{\chi}$ are known, then direct detection data from different detector materials and different energies can be used to solve for $\sqrt{\rho_{\chi}}f_{p,n}$ and $\sqrt{\rho_{\chi}}a_{p,n}$ by inverting equation \ref{rate}. If the halo velocity profile can be approximated, only $m_{\chi}$ is required to compute $\sqrt{\rho_{\chi}}f_{p,n}$ and $\sqrt{\rho_{\chi}}a_{p,n}$ with sufficient data. We can generally expect to have many more measurements than the minimum required to solve the system of equations once wimps are observed.\\
\indent Because the interaction parameters are constant, all linearly independent combinations of measurements used to solve for the scaled interaction parameters must agree, if the correct mass were used in the derivation. This motivates us to define a `kinematical consistency' function $\zeta(m'_{\chi})$,
\begin{equation*}
\zeta(m'_{\chi})\equiv\sqrt{\rho_{\chi}}\sum_{i\neq j}\left\{\left(a_p(i)-a_p(j)\right)^2+\left(a_n(i)-a_n(j)\right)^2+\mathrm{similar~terms}\right\},
\end{equation*}
which compares the values of $\sqrt{\rho_{\chi}}f_{p,n}$ or $\sqrt{\rho_{\chi}}a_{p,n}$ obtained using different independent subsets of the data---indexed by $i,j$---as a function of $m'_{\chi}$ used to invert the equations. It is obviously necessary that $\zeta(m'_{\chi})=0$ when $m'_{\chi}=m_{\chi}$\footnote{This may not be a sufficient condition, however; although, we have found no example where $\zeta(m'_{\chi})=0$ when $m'_{\chi}\neq m_{\chi}$.}.\\
\indent To determine the wimp mass, one varies $m'_{\chi}$ until $\zeta(m'_{\chi})=0$. We applied this test to some six thousand random, constrained MSSMs. For every single model tested, the correct mass was determined to near-arbitrary precision. Figure \ref{mass_scan} illustrates a typical plot of $\zeta(m'_{\chi})$. Notice that it has an extremely sharp minimum within a few GeV of the true wimp mass. It should be noted, however, that experimental uncertainties and resolutions were not considered during these calculations.

\vspace{-0.3cm}
\section{Neutralino Dark Matter and a Strict Lower Bound on $\rho_{\chi}$}

\indent Before we can compute $\rho_{\chi}$, the interaction parameters $f_{p,n}$ or $a_{p,n}$ must be `known.' These interaction parameters depend on very detailed knowledge of the particle physics of $\chi$. Let us consider the specific case in which the discovered particle is the neutralino. The interaction parameters will then depend on many details of the MSSM---these may not be known until well after dark matter particles have been discovered. Nevertheless, even with extremely limited knowledge of the MSSM, we can estimate the parameters using partial data, bounds, and constraints.\\
\indent  For example, we have found that given bounds on $\tan\beta$ and a lower bound on the lightest squark mass, $m_{\tilde{q}}$, there is a strict upper bound for the incoherent $\chi$-quark scattering parameters. In this case, it can be shown, that the magnitude of $a_{u}$ is strictly bounded by
\begin{align*}
\hspace{-2cm}a_u\leq &-\frac{g^2}{16m_W^2}(N_{\tilde{H}_1}^2-N_{\tilde{H}_2}^2)+\frac{g^2}{8}\frac{1}{(m_{\tilde{q}_{\ell}}^2-(m_{\chi}^2+m_u)^2}\left\{\raisebox{0.5cm}{$\!$}\frac{17}{18}\tan^2\theta_WN_{\tilde{B}}^2+\frac{1}{2}N_{\tilde{W}}^2\right.\\
&\left.+\frac{m_u^2}{m_W^2\sin^2\beta_{\ell}}N_{\tilde{H}_2}^2\right.+\frac{1}{3}\tan\theta_W|N_{\tilde{B}}||N_{\tilde{W}}|\cos(\alpha_{\tilde{W}})\\
&+\frac{m_u}{m_W\sin\beta_{\ell}}|N_{\tilde{W}}||N_{\tilde{H}_2}|\cos(\alpha_{\tilde{H}_2}-\alpha_{\tilde{W}})-\frac{m_u}{m_W\sin\beta_{\ell}}\tan\theta_W|N_{\tilde{B}}||N_{\tilde{H}_2}|\cos(\alpha_{\tilde{H}_2})\left.\raisebox{0.5cm}{$\!$}\right\},
\end{align*}
where $\alpha_{\tilde{H}_2}$, and $\alpha_{\tilde{W}}$ are the relative phases between $N_{\tilde{H}_2},N_{\tilde{W}}$ and $N_{\tilde{B}}$, respectively. This expression has six real unknowns. Notice that by the normalization of the neutralino wave function, the parameter space is compact. Therefore, $a_{u}$ can be {\it absolutely} maximized with respect to all six unknowns. It should be emphasized that this analysis is for the most general softly-broken MSSM; no {\it ad hoc} supersymmetry breaking scenarios---such as mSUGRA---were assumed.\\
\indent Because we can determine $\sqrt{\rho_{\chi}}a_{p,n}$, the strong upper bounds on $a_{p,n}$ directly translate into strong lower bounds on $\rho_{\chi}$. To test this idea, we considered some six thousand randomly generated, constrained MSSMs. For each of these models, upper bounds were calculated for $a_{p,n}$ assuming $10\%$ uncertainty in $\tan\beta$ and a lower bound on the lowest squark mass of either $200$ GeV or the actual mass of the lightest squark, whichever is less. The specific gauge content of the neutralino was taken to be known for each model for computational simplicity\footnote{If the gauge content of the neutralino was unknown, the interaction parameters could have been maximized with respect to these parameters as described earlier. In general, therefore, the upper bounds plotted are more restrictive than they would be in practice.}. Using the upper bounds for $a_{p,n}$, we obtain a lower bound on the local density $\rho_{\chi}$.\\
\begin{figure}[t]\vspace{-1cm}\centering\includegraphics[scale=0.182]{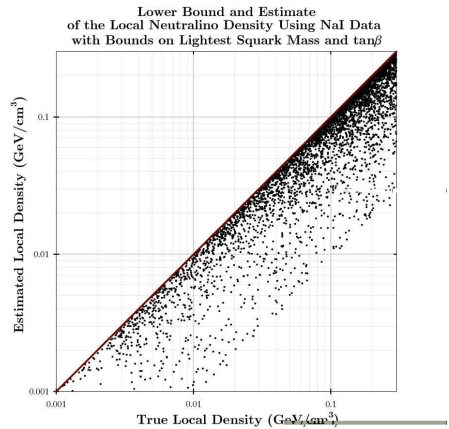}\caption{This plot compares the lower bound and estimate of the local denisty computed using the strong upper bound for $a_{p,n}$ to the true local density for each model.}\label{density_bound}\end{figure}
\indent Figure \ref{density_bound} illustrates the results of using this algorithm. Notice that the estimated local density is always strictly less than the true local density. Also, for many models the lower bound is not such a poor estimate.
\vspace{-0.3cm}\section{Conclusions}
\indent We have seen that, by itself, a discovery of dark matter particles in our galactic halo cannot solve the dark matter problem. However, combined with data from colliders to identify a particle and determine its interaction parameters, we can generally estimate its local density.\\
\indent We presented a robust, model independent method to determine the mass of a wimp using direct detector data alone. We have shown explicitly how one can determine or estimate $\rho_{\chi}$ in the case where $\chi$ is the neutralino.\\~\\
\indent Although the dark matter problem may not be solved immediately when wimps are discovered, there are clear and general ways to address their cosmological significance.\\

\noindent{\bf Acknowledgements}\\
\indent This research was done in collaboration with Gordon Kane of the University of Michigan and was supported by the Michigan Center for Theoretical Physics and the National Science Foundation's 2004 REU program.\\
\indent I would like to thank the organizers of the 5$^{\mathrm{th}}$ International Workshop on the Identification of Dark Mater for the wonderful program.

\vspace{-0.3cm}

\end{document}